\documentclass[prl,twocolumn,amsmath,amssymb,floatfix,superscriptaddress,nofootinbib,showpacs,preprintnumbers]{revtex4-1}
\usepackage{amsfonts, amssymb}
\usepackage{graphicx, epsfig,bm,epsf}
\usepackage{color,natbib}
\usepackage{amsbsy}
\usepackage{amsmath}
\usepackage{mathrsfs,url, dcolumn}

\newcommand{\be}{\begin{equation}}
\newcommand{\ee}{\end{equation}}
\newcommand{\bea}{\begin{eqnarray}}
\newcommand{\eea}{\end{eqnarray}}

\def\fun#1#2{\lower3.6pt\vbox{\baselineskip0pt\lineskip.9pt
        \ialign{$\mathsurround=0pt#1\hfill##\hfil$\crcr#2\crcr\sim\crcr}}}

\renewcommand\({\left(}
\renewcommand\){\right)}
\renewcommand\[{\left[}
\renewcommand\]{\right]}

\newcommand\lsim{\mathrel{\rlap{\lower4pt\hbox{\hskip1pt$\sim$}}
    \raise1pt\hbox{$<$}}}
\newcommand\gsim{\mathrel{\rlap{\lower4pt\hbox{\hskip1pt$\sim$}}
    \raise1pt\hbox{$>$}}}

\newcommand{\sPdd}{\ensuremath{\mathscr{P}_{\delta\delta}}}
\newcommand{\sPxd}{\ensuremath{\mathscr{P}_{x\delta}}}
\newcommand{\sPxx}{\ensuremath{\mathscr{P}_{xx}}}
\newcommand{\xH}{\ensuremath{\bar{x}_{\rm H}}}
\newcommand{\fnl}{\ensuremath{f_{\rm NL}}}
\newcommand{\PDT}{\ensuremath{P_{\Delta T}}}

\newcommand{\Pdd}{\ensuremath{P_{\delta\delta}}}

\def\dslash{\not{\hbox{\kern-2pt $\partial$}}}
\def\Dslash{\not{\hbox{\kern-4pt $D$}}}
\def\Oslash{\not{\hbox{\kern-4pt $O$}}}
\def\Qslash{\not{\hbox{\kern-4pt $Q$}}}
\def\pslash{\not{\hbox{\kern-2.3pt $p$}}}
\def\kslash{\not{\hbox{\kern-2.3pt $k$}}}
\def\qslash{\not{\hbox{\kern-2.3pt $q$}}}

 \newtoks\slashfraction
 \slashfraction={.13}
 \def\slash#1{\setbox0\hbox{$ #1 $}
 \setbox0\hbox to \the\slashfraction\wd0{\hss \box0}/\box0 }

\def\ee{\end{equation}}
\def\be{\begin{equation}}

\newcommand\bfk{{\bf k}}

\newcommand\bfu{{\bf u}}

\usepackage{multirow}

\newcommand{\newsec}[1]{\vspace{0.25cm}\noindent{\bf \emph{#1}}.\hspace{0.25cm}}

\begin{document}
\setlength{\unitlength}{1mm}

\title{Primordial non-Gaussianity from the 21 cm Power Spectrum \\
during the Epoch of Reionization}

\author{Shahab Joudaki}
\affiliation{Center for Cosmology, Dept. of Physics \& Astronomy, University of California, Irvine, CA 92697}

\author{Olivier Dor\'e}
\affiliation{Jet Propulsion Laboratory, California Institute of Technology, Pasadena, CA 91109}
\affiliation{California Institute of Technology, Pasadena, CA 91125}

\author{Luis Ferramacho}
\affiliation{CNRS, IRAP, 14 Avenue Edouard Belin, F-31400, Toulouse, France}
\affiliation{Universit\'e de Toulouse, UPS-OMP, IRAP, Toulouse, France}

\author{Manoj Kaplinghat}
\affiliation{Center for Cosmology, Dept. of Physics \& Astronomy, University of California, Irvine, CA 92697}

\author{Mario G. Santos}
\affiliation{CENTRA, Departamento de F\'{i}sica, Instituto Superior Tecnico, 1049-001 Lisboa, Portugal}

\date{\today}

\begin{abstract}
Primordial non-Gaussianity is a crucial test of inflationary
cosmology. We consider the impact of non-Gaussianity on the
ionization power spectrum from 21 cm emission at the epoch of
reionization. We focus on the power spectrum on large scales at
redshifts of 7 to 8 and explore the expected constraint on the local
non-Gaussianity parameter $f_{\rm NL}$ for current and 
next-generation 21 cm experiments. We show that experiments such as
SKA and MWA could measure $f_{\rm NL}$ values of order 10. This can be improved
by an order of magnitude with a fast-Fourier transform telescope like Omniscope. 
\end{abstract}
\bigskip

\maketitle

\newsec{Introduction}
An inflationary epoch in the early universe~\cite{Guth,Linde:1981mu} has been
established as a solution to the cosmological  
horizon and flatness problems over the past three decades, most
recently through high-precision measurements of the cosmic microwave
background (CMB) by the Wilkinson Microwave Anisotropy Probe
(WMAP)~\cite{Spergel:2003cb}. The inflationary hypothesis predicts an
epoch of exponential growth lasting at least 60 e-folds resulting in
almost Gaussian scale-invariant density
perturbations~\cite{Bartolo:2004if}.   

A powerful mechanism to distinguish between inflation models is the
amplitude and scale dependence of mild non-Gaussianity in
perturbations of the primordial density field. 
Canonical single field inflation models predict primordial non-Gaussianity (bispectrum) of
the local form $|\fnl| \ll 1$~\cite{Maldacena:2002vr,Acquaviva:2002ud}, while evolution after
inflation generates non-local bispectrum with effective $\fnl = {\cal O}(1)$ \cite{Verde2000,Liguori:2005rj,Smith:2006ud}. 
The best current constraints of $\pm 25$ on local $\fnl$~\cite{Smith:2009jr,
Smidt:2009ir} are from WMAP data. A future measurement of $\fnl={\cal
O}(1)$ could reveal the existence of physics beyond the standard
single field slow-roll inflationary scenario.

We show that radio interferometric
probes~\cite{lofar,mwa,ska,omniscope} of 21~cm emission from spin-flip
transitions of neutral hydrogen at the epoch of reionization
(EoR)~\cite{Furlanetto:2006jb} can result in constraints on $\fnl$ at
the same level as Planck~\cite{planckbb}, and less than unity in the
most optimistic experimental proposal.   
Previous studies have explored primordial non-Gaussianity in the
bispectrum of ideal 21~cm experiments prior to the
EoR~\cite{Cooray:2006km, Pillepich:2006fj}. In this work, we consider 
scale dependent bias in the power spectrum of ionized hydrogen
resulting from departures from Gaussian initial
conditions~\cite{dalaldore, verde}. 
Our constraints from 21~cm emission do not require an ionization-clean cosmology, 
i.e., a priori knowledge of the spectrum of fluctuations in the ionized fraction.

The rest of the letter is arranged as follows. We first quantify the
influence of non-Gaussianity of the local form on the 21 cm power
spectrum, and then test this via numerical simulations of the
ionization distribution. We review the assumed noise properties of 
LOFAR~\cite{lofar}, MWA~\cite{mwa}, SKA~\cite{ska}, and
Omniscope~\cite{omniscope}, and forecast constraints on $\fnl$ based on a Fisher matrix analysis. For these forecasts,
we fix the parameters of our fiducial flat $\Lambda$CDM model to agree
with WMAP7~\cite{Komatsu:2010fb}. 

\begin{figure}[!t]
\begin{center}
\vspace{-0.8em}
\centerline{\includegraphics[scale=0.46]{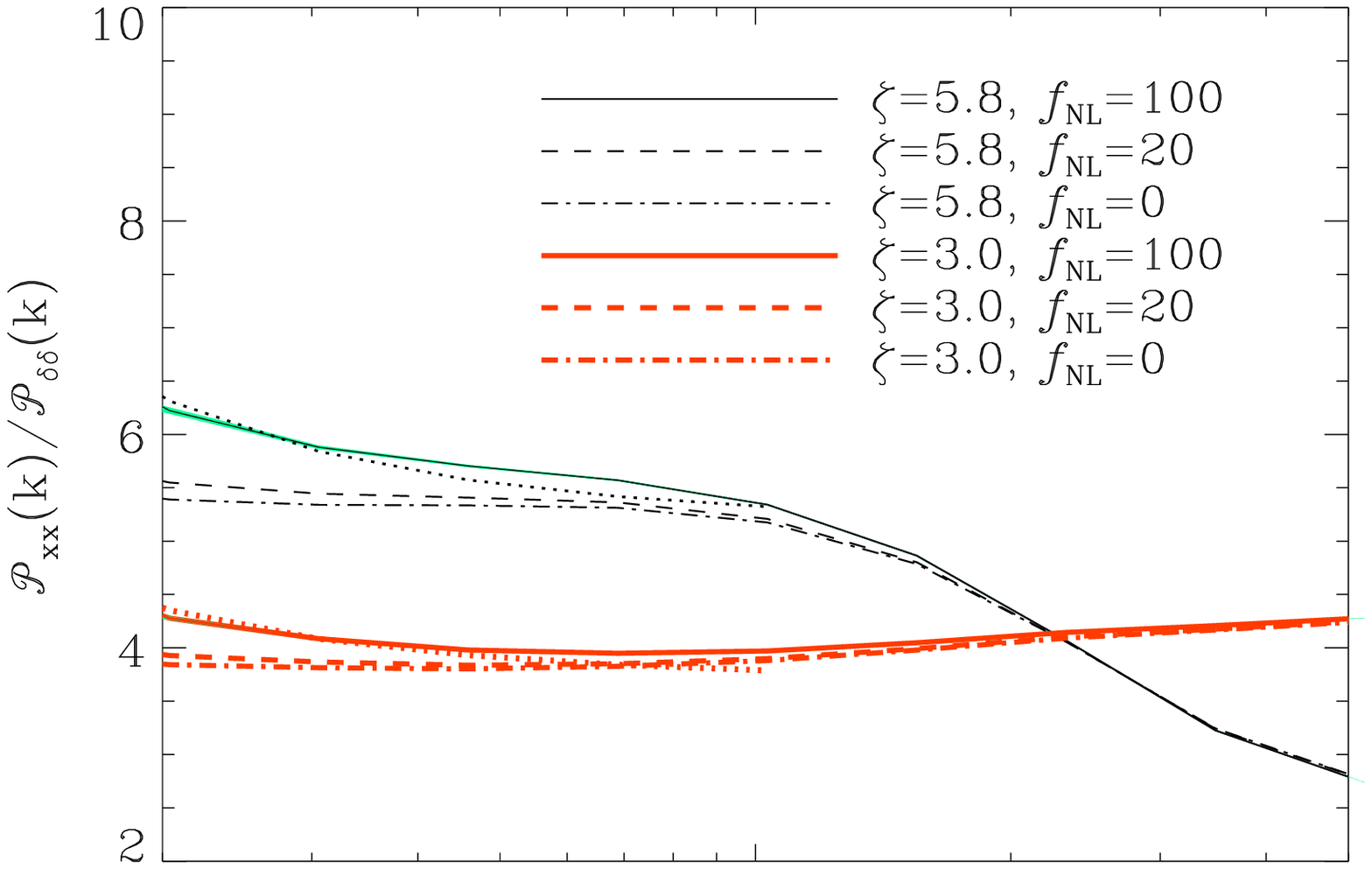}}
\vspace{-3.2em}
\centerline{\includegraphics[scale=0.46]{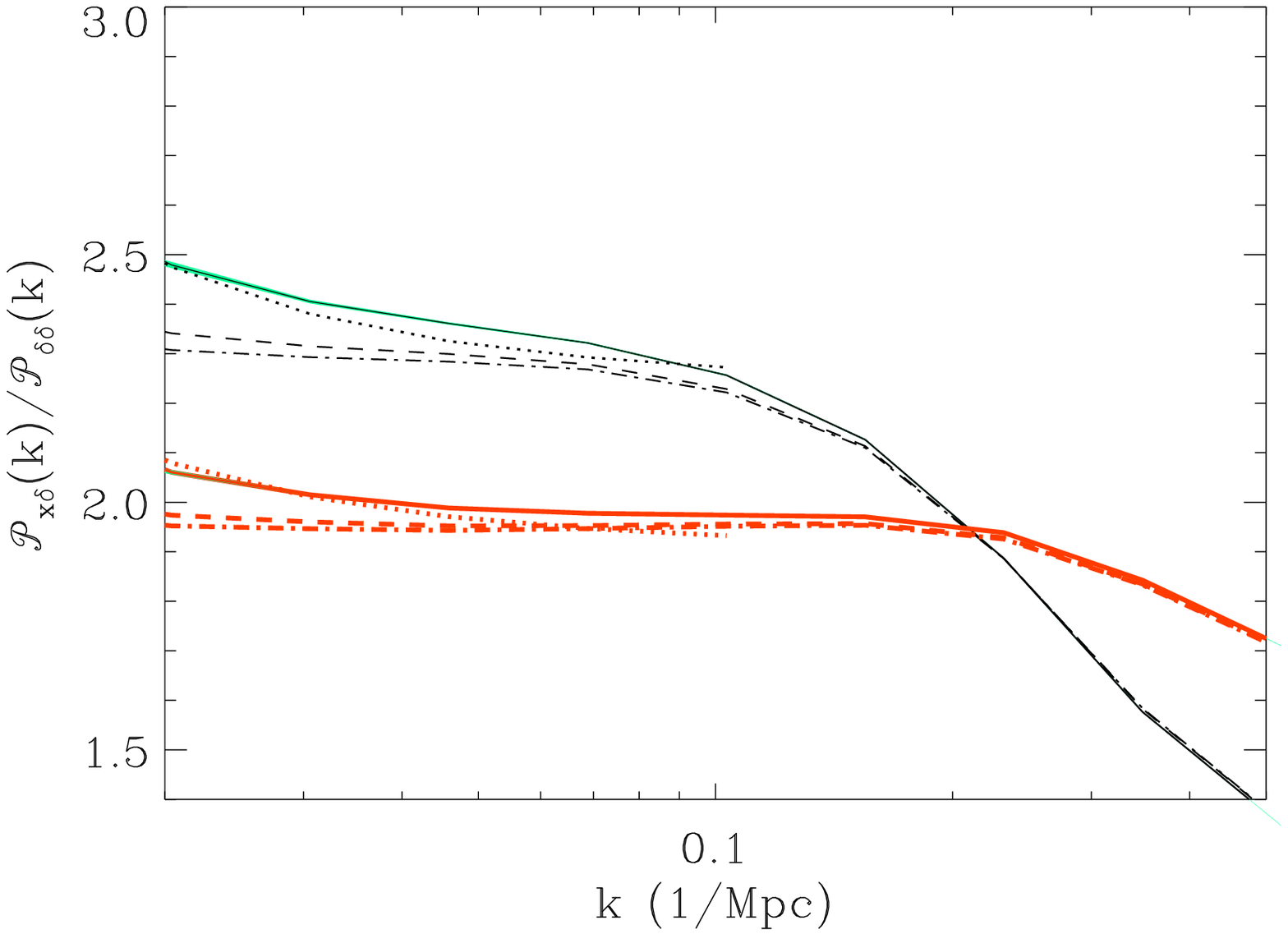}}
\end{center}
\vspace{-3em}
\caption{Ionization power spectra 
  with non-Gaussianity of the local form from numerical simulations. 
We show $\fnl = \(0,20,100\)$ (dot-dashed, dashed, solid) for
efficiency $\zeta = \(5.8, 3.0\)$ (thin black, thick red) at $z =
7.5$, where $\xH = \(0.50, 0.75\)$. For $\fnl=100$ cases, sample variance from simulations
is in form of green bands about the mean, and analytical fits corresponding to ${\bar \delta}_B = 1$ are in dotted lines.}
\label{figurett}
\end{figure}

\newsec{Effect of Non-Gaussianity on the 21 cm Power Spectrum}
We decompose the 21 cm power spectrum at redshift $z$ in terms of its angular dependence~\cite{Barkana:2004zy}, given by $\mu=\hat{\bf k}\cdot\hat{{\bf n}} = \cos(\theta)$, where $\theta$ is the angle between wavevector ${\bf k}$ and line of sight (LOS) vector ${\bf n}$:
\bea
& & P_{{\Delta}T}({\bf k},z) = \sPdd(k,z) - 2\sPxd(k,z) + \sPxx(k,z) \nonumber\\ 
& & \quad + 2\left[\sPdd(k,z) - \sPxd(k,z)\right]\mu^2 + \sPdd(k,z)\mu^4.
\eea
We define $\sPdd \equiv \tilde{T}_b^2 \xH^2 \Pdd$~\cite{maoteg}, where $\Pdd$ is the linear matter power spectrum, numerically obtained from a modified version of CAMB~\cite{LCL}, $\xH$ is the mean neutral fraction of hydrogen such that the ionized fraction ${\bar{x}_{\rm i}} \equiv 1-\xH$, and $\tilde{T_b}(z=7.5) \simeq 0.026~{\rm K}$ is the spatially averaged brightness temperature.
We consider only large enough scales ($k<0.15/{\rm Mpc}$) such that the ionization power spectrum $\sPxx \simeq b_x^2 \sPdd$ and the ionization-density cross spectrum $\sPxd \simeq b_x \sPdd$, where $b_x$ is the bias of ionized regions.
Our numerical simulations in Fig.~\ref{figurett} show that this is an excellent approximation.

We define ${\bf u}$ as the Fourier dual of ${\bf \Theta} \equiv \theta_i \hat{e}_i +\theta_j \hat{e}_j + \Delta f \hat{e}_k$, where 
$\theta_i$ and $\theta_j$ encode the angular location on the 2D sky, and $\Delta f$ measures the difference in frequency. The 21 cm power spectrum is extended to ${\bf u}$-space in which measurements are made: 
\be
\PDT(\bfu,z)= \PDT(\bfk,z)/{\left(\chi^2(z) y(z)\right)} ,
\label{ktou}
\ee
where $\chi(z)$ is the comoving distance to a given redshift, $y(z) = \lambda_{21}(1+z)^2/H(z)$ 
translates between intervals in frequency and distance, 
and $\lambda_{21} = \lambda(z)/(1+z) = 0.21~{\rm m}$.
We convert between $u$ and $k$ spaces via $u_\perp = \chi(z) k_\perp = 2\pi L/\lambda(z)$, where $L$ is the baseline, and $u_\parallel = y(z) k_\parallel$.

Given non-Gaussianity of the local form, Bardeen's gauge invariant
potential field $\Phi$ is related to a pure gaussian random field
$\phi$ at nonlinear order~\cite{Salopek:1990jq,Verde2000}: 
\be
\Phi_{\rm NG}({\bf x})=\phi({\bf x})+\fnl\left(\phi^2({\bf x})-\langle \phi^2 \rangle\right) .
\label{local_nongauss}
\ee  
In the high-peaks formalism $\fnl$ influences biased tracers of the
underlying matter distribution as a scale dependent correction to the
large scale bias~\cite{dalaldore,verde}. This enters as $\sPxd/\sPdd =
b_x + {\Delta}b_x$, $\sPxx/\sPdd = (b_x + {\Delta}b_x)^2$, with
\be
{\Delta}b_x(k,z) = {3(b_x-1)\fnl\Omega_mH_0^2{\bar \delta}_B} / {\left(D(z)k^2T(k)\right)} ,
\label{dbx}
\ee
where $H_0$ is the Hubble constant, $\Omega_m$ is the present density
parameter of matter, $D(z)$ is the linear growth function of density
perturbations, and $T(k)$ is the transfer function relating
present and primordial power spectra. 
The quantity ${\bar \delta}_B$ is the average critical collapse density of HII regions~\cite{Furlanetto:2004nh}. 
We leave the bias $b_x$ as a free parameter, although $b_x$, $\delta_B$, and $\xH$ would all be related in a given model of
reionization. The scale-dependence of the bias in $\Delta b_x$ is clearly evident in
the ionization spectra from our simulations in
Fig.~\ref{figurett}. We find that ${\bar \delta}_B \sim 1$ fits the
large-scale $\fnl$ induced rise to the ionization spectrum. 

\begin{figure}[!t]
\begin{center}
\vspace{-0.8em}
\centerline{\includegraphics[scale=0.46]{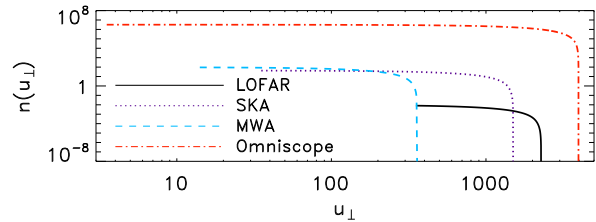}}
\end{center}
\vspace{-3.3em}
\caption{Number density of baselines for LOFAR (solid), SKA (dotted), MWA (dashed), and Omniscope (dot-dashed).}
\label{figurtva}
\end{figure}

\begin{table}[!t]\footnotesize
\begin{center}
\begin{tabular}{lcccccccccccccc|c}
\hline
\hline
Experiment & $N_{\rm ant}$ & $L_{\rm min}$ (m) & FOV (deg$^2$) & $A_e ({\rm m}^2)$ \cr
\hline
LOFAR &      32 &       100 &      $2 \times \pi 2.4^2$ &       590 \cr
MWA &      500 &       4.0 &      $\pi 16^2$ &       13  \cr
SKA &      1400 &       10 &      $\pi 8.6^2$ &       45 \cr
Omniscope & $10^6$ & 1.0 &     $2.1 \times 10^4$ &      1.0 \cr
\hline
\hline
\end{tabular}
\caption{Experimental specifications for the telescopes. The antenna number only account for those inside the nucleus and core (e.g. for SKA we use 1400 of a total 7000 antennae). The system temperature $T_{\rm sys} = 390~{\rm K}$, bandwidth is 6 MHz, observation time is 4000 hours, and effective area at $z=7.5$.}
\label{tableone}
\end{center}
\end{table}

\newsec{Numerical Simulations with Non-Gaussian Initial Conditions}
We perform simulations of the ionization distribution during the EoR for $\fnl=(0,20,100)$ and ionization efficiency $\zeta=(3.0,5.8)$, in a box of comoving length $3000~{\rm Mpc}$, with a modified version of SimFast21~\cite{simfast21,simfastlink}.
The initial matter density field is computed from the Poisson equation with non-Gaussian gravitational
potential $\Phi_{\rm NG}({\bf k})$. 
We show the spectra from these simulations in Fig~\ref{figurett}, from which ${\bar \delta}_B \sim 1$.

We compare this result to the theoretical prediction. 
The critical density for collapse of an ionized region of mass $m$ is obtained from the collapse fraction $f_{\rm coll}$~\cite{Furlanetto:2004nh}:
\be
\delta_B(m, z) = \delta_c -{K}(\zeta)\sqrt{2\left(\sigma^2(m_{\rm min},z) - \sigma^2(m,z)\right)},
\label{massdep}
\ee
where $\delta_c \approx 1.68$ is the critical collapse density of matter, $\sigma^2(m,z)$ is the variance of the density fluctuations, and $m_{\rm min}$ corresponds to a virial temperature of $10^4~{\rm K}$. Moreover, ${K}(\zeta) = {\rm erf}^{-1}(1 - \zeta^{-1})$, where $\zeta = m_{\rm ion}/m_{\rm gal}$ is the ionization efficiency~\cite{Furlanetto:2004nh}. We evaluate ${\bar \delta}_B$ as an average over the fraction of space filled by HII bubbles as in Ref.~\cite{Furlanetto:2004nh}.
Given this prescription, we find ${\bar \delta}_B = 1.1$ (less than
$\delta_c$ as $\zeta>1$), matching the simulation results well. 
This becomes ${\bar \delta}_B = 1.2$ if we only average over the mass
function. For simplicity, we fix ${\bar \delta}_B = 1$.

As noted earlier, the bias $b_x$, collapse threshold $\delta_B$, and  $\xH$
are expected to be interrelated in a given reionization scenario.  This is evident in
Fig.~\ref{figurett}, where we see that a factor of 2 change in $\zeta$
changes the bias by about 15\%. This change is subdominant to the
impact of $\xH$ (linear function of $\zeta$) on the 21 cm
power spectrum. In a more optimistic scenario, one could envision
constraining $\xH$ (or $\zeta$) together with $\fnl$ without
$b_x$ as a free parameter. We also considered the
impact of variations in $\xH$ and $\fnl$ on ${\bar
  \delta}_B$. Changing $\xH$ by a factor of two 
only affects ${\bar
  \delta}_B$ by 8\% given $(1-\xH) = \zeta f_{\rm
  coll}$. Nonzero $\fnl$ skews 
$\delta_B(m,z)$ through its influence on $f_{\rm
  coll}$. Using the results of
Ref.~\cite{Matarrese:2000iz}, we estimate 
${\bar \delta}_B$ is only perturbed by 4\% even for $\fnl=100$. This is because the sensitivity to $\fnl$ increases with mass, while the mass scales that contribute a majority of the integral over the mass
function lie within an order of magnitude of the minimum halo mass. 

\newsec{21 cm Noise Power Spectrum}
The noise power spectrum of 21 cm fluctuations is expressed as~\cite{maoteg,mcquinn}
\be
P_N(u_\perp,z)=\left({\lambda^2(z) T_{\rm sys}(z)}/{A_e(z)}\right)^2/\left({t_0 n(u_\perp)}\right) ,
\ee
where the sky-dominated system temperature $T_{\rm sys} \simeq 280\left((1+z)/7.4\right)^{2.3}\rm{K}$~\cite{Wyithe:2007if}, $t_0$ is the total observation time, and $A_e(z) \propto \lambda^2(z)$ is the effective collecting area (listed in Table~\ref{tableone}). Here, $n(u_\perp)$ encodes the number density of baselines shown in Fig.~\ref{figurtva}, computed as the autocorrelation of the array density for each of the surveys. 

The array distributions are composed of a nucleus with full coverage
fraction and a core with power law $r^{-2}$. The nucleus radius is 
$R_n = \sqrt{\eta N_{\rm ant}/(\rho_0\pi)}$, 
where $\rho_0$ is the 2D array density of the nucleus, and $N_{\rm ant}$ is the number of antennae of each experiment (see Table~\ref{tableone}). The core radius is by construction $R_c = R_n \exp((1-\eta)/(2\eta))$~\cite{maoteg}. The most optimal choice of $\eta$ for constraints on $\fnl$ depends on the particular experiment and bandwidth $B$ considered, but for comparison with prospective constraints on other cosmological parameters in Table V of Ref.~\cite{maoteg}, we choose $\eta=0.8$ for $\[\rm{LOFAR, MWA, SKA}\]$, whereas all of Omniscope's antennae lie in the nucleus. 

\begin{figure}[!t]
\begin{center}
\vspace{-0.8em}
\centerline{\includegraphics[scale=0.46]{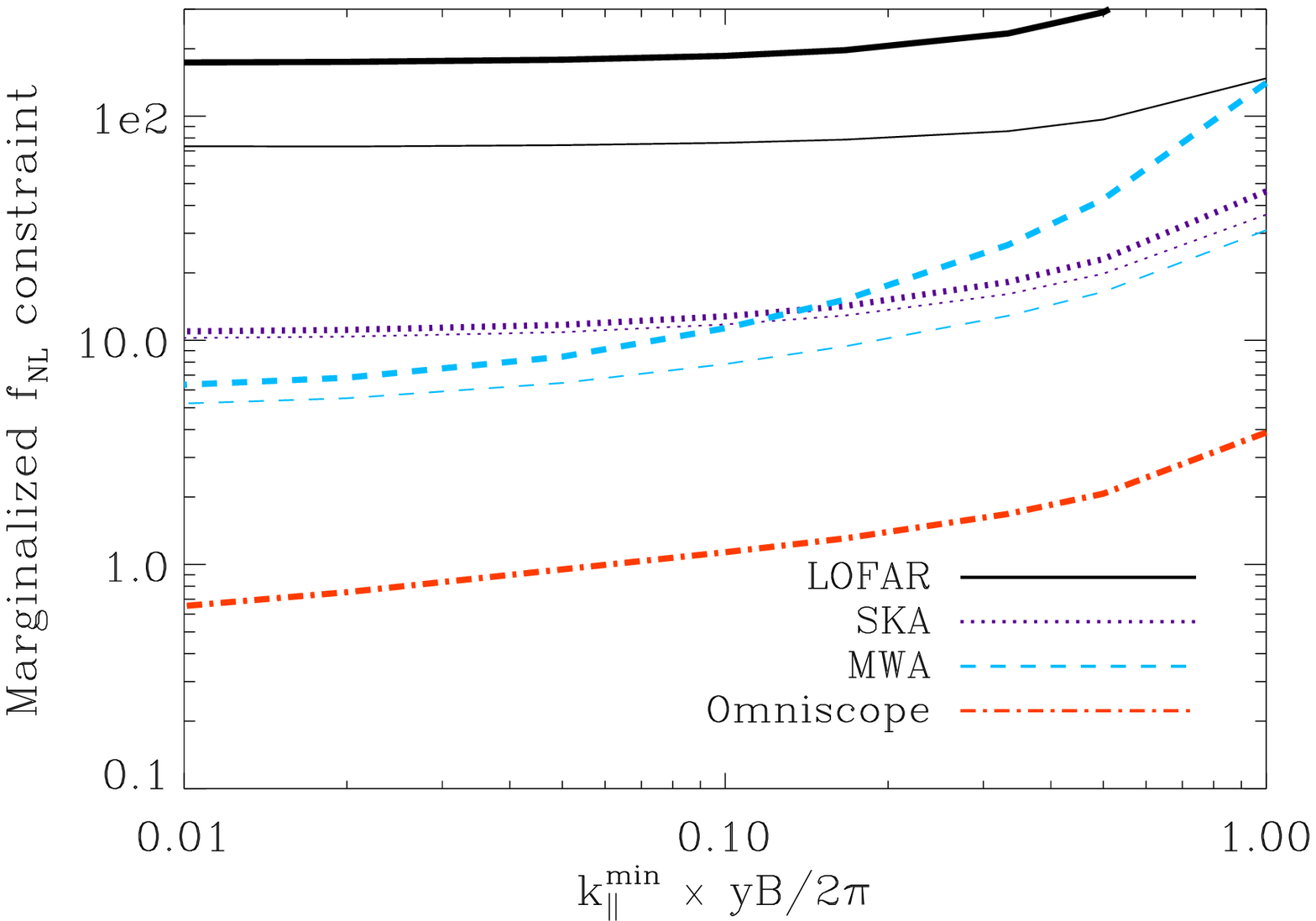}}
\vspace{-0.5em}
\centerline{\includegraphics[scale=0.46]{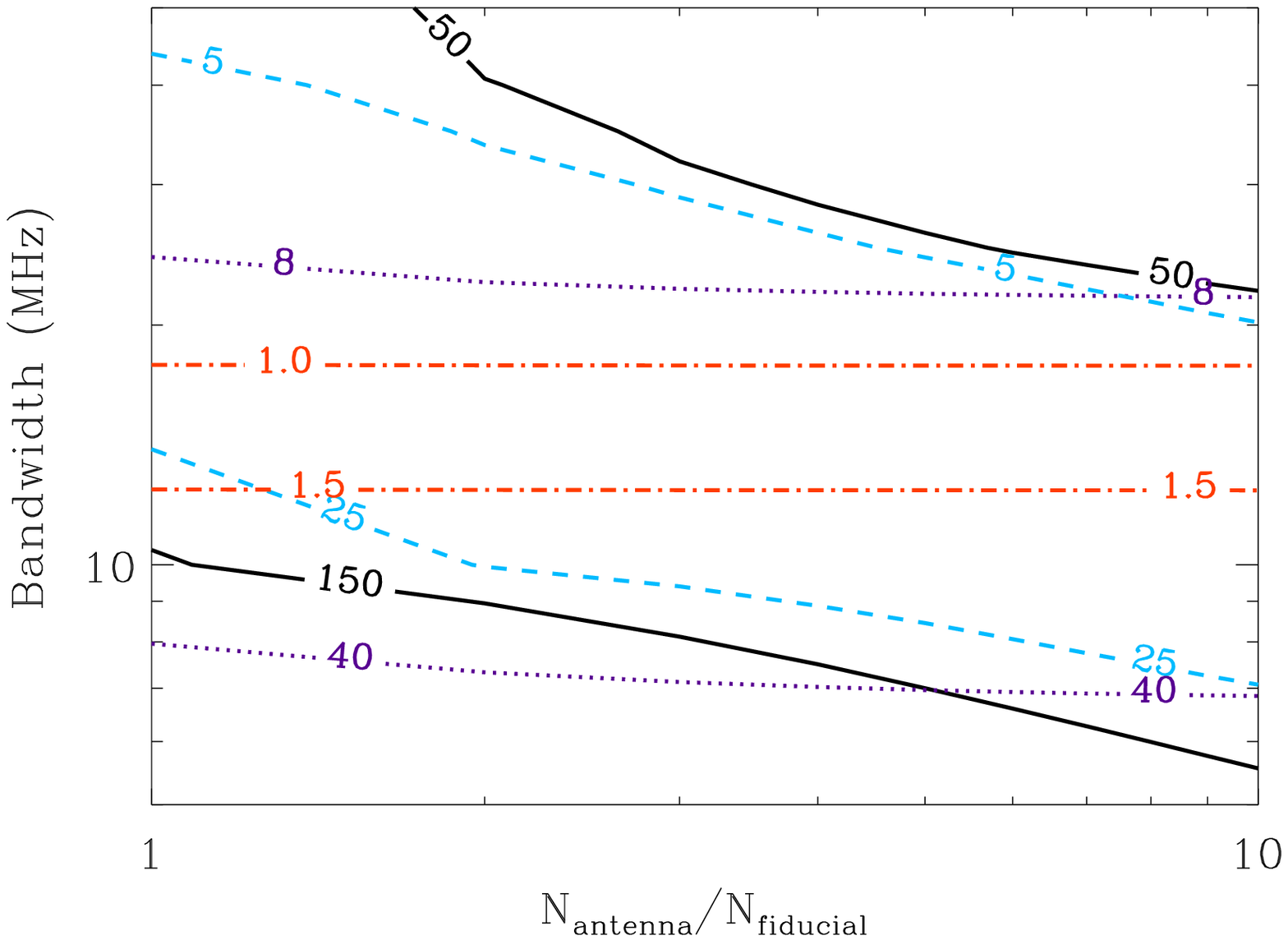}}
\end{center}
\vspace{-3em}
\caption{{\it Top}: Marginalized $\fnl$ constraints for cases with noise (thick) and without noise (thin), which overlap for Omniscope. We consider a bandwidth of 6 MHz, but assume foregrounds can be removed on scales larger than $k_\parallel = 2\pi/(yB)$. 
{\it Bottom}: Marginalized $\fnl$ constraints as function of bandwidth and number of antennae.
The bandwidth limits the number of modes and largest scale probed along the LOS (via the survey volume $V \propto B$ and $k_\parallel^{\rm min} \propto 1/B$), whereas a larger number of antennae for fixed array density increases the survey resolution and number of perpendicular modes (via $n(u_\perp)$, on large scales $\propto N_{\rm ant}$, and $u_\perp^{\rm max} \propto \sqrt{N_{\rm ant}}$).
The color coding is the same as for the top panel.}
\label{figurtre}
\end{figure}

We assume residual foregrounds can be ignored beyond $k_\parallel \geq 2\pi/(yB)$~\cite{mcquinn}, but also consider the case where foregrounds can be removed on larger scales (Fig~\ref{figurtre}).

\newsec{Fisher Matrix Forecasts}
We evaluate the prospective constraints on $\fnl$ from the 21 cm power spectrum at the EoR via the
Fisher matrix formalism. The summation involves pixels in $(k_\perp,k_\parallel)$ of thickness $(\epsilon_\perp,\epsilon_\parallel) = (\Delta{k_\perp}/k_\perp,\Delta{k_\parallel}/k_\parallel) = (0.1, 0.1)$:
\be 
{\bf F}_{ab} = \sum_{\rm pixels} \frac{1}{[\delta \PDT({\bf u})]^2}\left(\frac{\partial \PDT({\bf u})}{\partial p_a}\right) \left(\frac{\partial \PDT({\bf u})}{\partial p_b}\right)\,.
\ee
We have verified that our forecasts are robust to variations in the
step sizes of parameter space and ${\bf k}$-space. 
The measurement error consists of the sum of the sample variance and
thermal detector noise over half-space~\cite{mcquinn}: 
\be
\delta\PDT({\bf u}) =  {\left(\PDT({\bf u})+P_N(u_\perp)\right)}/\sqrt{N_m} .
\ee
The number of modes falling in each pixel is given by $N_m = 2\pi k_\perp
\Delta k_\perp \Delta k_\parallel {\rm V(z)}/(2\pi)^3$,  
such that the volume sampled $V(z) = \chi^2 y B \times {\rm FOV}$,
where FOV denotes the field of view of the telescope (often equal to
$\lambda^2/A_e$). 

For a single redshift bin at $z = 7.5$, we fiducially let $b_x = 2.3$ and $\xH = 0.5$. 
The bandwidth $B = 6~{\rm MHz}$ limits $k_\parallel^{\rm min} = 2\pi/(yB) \gsim 0.063/{\rm Mpc}$~\cite{mcquinn}, and nonlinearities force ${k_\parallel^{\rm max}}\sim{2/{\rm Mpc}}$. The ranges in $k_\perp$ at the central redshift are $\left[0.039, 0.25\right]/{\rm Mpc}$ for LOFAR, $\left[0.0016, 0.040\right]/{\rm Mpc}$ for MWA, $\left[0.0039, 0.17\right]/{\rm Mpc}$ for SKA, and $\left[3.9\times10^{-4}, 0.44\right]/{\rm Mpc}$ for Omniscope. However, due to our narrow focus on $\fnl$ at the largest scales in which the $1/k^2$ boost becomes significant, in practice, we only keep modes up to $k^{\rm max} = 0.15/{\rm Mpc}$.

\newsec{Results}
In quantifying our constraints on $\fnl$, we fix the underlying cosmology. By only considering large enough scales for which the ratio of the ionization and matter spectra is constant in a universe without non-Gaussianity, the free parameters in a single redshift bin at $\xH = 0.5$ are limited to $(\fnl, b_x, \xH)$.
With Planck priors on the standard cosmological
parameters~\cite{planckbb}, in particular the matter power
spectrum normalization $\Delta_R^2$, cold dark matter density
${\Omega_c}h^2$, spectral index $n_s$, and its running ${\rm
  d}{n_s}/{\rm d}\ln{k}$, we find the $\fnl$ constraints from [LOFAR,
MWA, SKA] are robust to the assumption of a fixed cosmology at the
10\% level, while the same level of robustness for Omniscope is
achieved after including its constraints on $\left(n_s, {\rm
    d}{n_s}/{\rm d}\ln{k}, {\Omega_c}h^2\right)$ from small
scales. The constraints on $\fnl$ will depend on the fiducial $b_x$, but we do
not explore this issue here.

Fig.~\ref{figurtre} (top) shows $\fnl$ constrained as function of the minimum LOS wavenumber, limited by the experimental ability to remove foregrounds. Imposing $k_\parallel^{\rm min} = 2\pi/(yB) = 0.063/{\rm Mpc}$~\cite{mcquinn}, we find the constraints for [LOFAR, MWA, SKA, Omniscope] are equal to $\sigma(\fnl) = \[700, 100, 50, 4\]$, which reduces to $\sigma(\fnl) = \[100, 30, 40, 4\]$ when instrumental noise is neglected. These constraints improve for telescopes with increased ability to probe larger LOS scales. When arbitrarily large scales along the LOS can be probed, we find $\sigma(\fnl) = \[200, 6, 10, 0.6\]$, which reduces to $\sigma(\fnl) = \[70, 5, 10, 0.6\]$ when noise is neglected. The constraints plateau for $k_\parallel^{\rm min} \rightarrow 0$ due to the nonzero $k_\perp^{\rm min}$ set by the minimum experiment baseline. 
As $k_\parallel^{\rm min}$ decreases, our assumed MWA configuration becomes somewhat better than the SKA configuration in constraining $\fnl$ due to its smaller minimum baseline, allowing larger scales to be probed by the telescope.

In Fig.~\ref{figurtre} (bottom), we consider a minimum LOS scale set by $k_\parallel^{\rm min} = 2\pi/(yB)$, but allow an order of magnitude variation in bandwidth and telescope antenna number. The bandwidth is inversely proportional to the minimum LOS wavenumber and linearly increases the volume probed, whereas larger number of antennae for fixed array density increases the maximum baseline as $\sqrt{N_{\rm ant}}$ and linearly boosts the baseline density (thereby decreasing the noise). The contours show increased bandwidth is more powerful in the search for $\fnl$, in particular for SKA and Omniscope that have small instrumental noise. This is because their signal-to-noise is already close to the cosmic variance limit, and our power spectrum cutoff at $k=0.15/{\rm Mpc}$ makes us insensitive to the increasing number of small scale modes. Extending the considered modes to scales of $k=2/{\rm Mpc}$ (incorporating modeling of the exponential tail with very strong priors on the new free parameters) improves the constraints by up to factor of 2 for the different experimental configurations.

We have also considered the case where the bias and ionization
fraction are fixed. In this scenario, the $\fnl$ constraints improve
by a factor of 1.5 up to a factor of 10 for the various cases and experiments
considered. For the fiducial configurations alone, the $\fnl$
constraints improve by factors of 2 (MWA) to 3 (LOFAR, SKA, Omniscope)
when fixing the bias to be a function of the ionization fraction. 
When only information from scales larger
than $k^{\rm max} = 0.10/{\rm Mpc}$ is available (compared to $0.15/{\rm
  Mpc}$ assumed throughout the paper), the constraint on $\fnl$ degrades by up to
a factor of 2 when marginalizing over $b_x$ and $\xH$, and by
up to $30\%$ when $b_x$ and $\xH$ are fixed. 

\newsec{Conclusions}
The search for a signature of primordial non-Gaussianity is a key test
of inflationary theories. Large values for the non-Gaussianity
parameter, $\fnl\gg 1$, will rule out standard single field
inflationary models. We have considered the impact of primordial
non-Gaussianity on the ionization power spectrum from 21 cm emission
at the epoch of reionization, which provides an alternative approach to constrain $\fnl$ relative to the cosmic microwave background and large-scale structure. 
We find that $\fnl$ can be constrained to an accuracy of order 10 with future 21 cm telescopes like SKA and
MWA. This improves by an order of magnitude for a fast-Fourier 
transform telescope like Omniscope, thereby opening a new window to
inflationary physics.

\smallskip
{\it Acknowledgements:} We thank A. Amblard, Y. Mao,
G. Martinez, M. McQuinn, J. Smidt, and E. Tollerud for useful
discussions. MGS acknowledges support by FCT
under grant PTDC/FIS/100170/2008. MK acknowledges support by NSF under
grant NSF 0855462 at UCI. Part of the research described in this letter was carried 
out at JPL, Caltech, under contract with NASA.


\begin{thebibliography}{99}
\frenchspacing

\bibitem{Guth}
  A.~H.~Guth,
  Phys.\ Rev.\  D {\bf 23}, 347 (1981).

\bibitem{Linde:1981mu}
  A.~D.~Linde,
  Phys.\ Lett.\  B {\bf 108}, 389 (1982).

\bibitem{Spergel:2003cb}
  D.~N.~Spergel {\it et al.}, 
  Astrophys.\ J.\ Suppl.\  {\bf 148}, 175 (2003).

\bibitem{Bartolo:2004if}
  N.~Bartolo, {\it et al.},
  Phys.\ Rept.\  {\bf 402}, 103 (2004).

\bibitem{Maldacena:2002vr}
  J.~M.~Maldacena,
  JHEP {\bf 0305}, 013 (2003).

\bibitem{Acquaviva:2002ud}
  V.~Acquaviva, {\it et al.},
  Nucl.\ Phys.\  B {\bf 667}, 119 (2003).

\bibitem{Verde2000}
  L.~Verde, {\it et al.},
  MNRAS {\bf 313}, 141 (2000).

\bibitem{Liguori:2005rj}
  M.~Liguori, {\it et al.},
  Phys.\ Rev.\  D {\bf 73}, 043505 (2006).

\bibitem{Smith:2006ud}
  K.~M.~Smith \& M.~Zaldarriaga,
  arXiv:0612571.

\bibitem{Smith:2009jr}
  K.~M.~Smith, {\it et al.},
  JCAP {\bf 0909}, 006 (2009).

\bibitem{Smidt:2009ir}
  J.~Smidt, {\it et al.},
  Phys.\ Rev.\  D {\bf 80}, 123005 (2009).
  
\bibitem{lofar}       
  http://www.lofar.org
 
\bibitem{ska}
  http://www.skatelescope.org

\bibitem{mwa}
  http://www.mwatelescope.org

\bibitem{omniscope}
  M.~Tegmark \& M.~Zaldarriaga,
  PRD {\bf 82}, 103501 (2010).
    
\bibitem{Furlanetto:2006jb}
  S.~Furlanetto, {\it et al.},
  Phys.\ Rept.\  {\bf 433}, 181 (2006).
  
\bibitem{planckbb}
  G.~Efstathiou, {\it et al.}, ESA-SCI 1 (2005).
     
\bibitem{Cooray:2006km}
  A.~Cooray,
  Phys.\ Rev.\ Lett.\  {\bf 97}, 261301 (2006).
  
\bibitem{Pillepich:2006fj}
  A.~Pillepich, {\it et al.},
  Astrophys.\ J.\  {\bf 662}, 1 (2007).
  
\bibitem{dalaldore}
  N.~Dalal, {\it et al.},
  Phys.\ Rev.\  D {\bf 77}, 123514 (2008).

\bibitem{verde}
  S.~Matarrese \& L.~Verde,
  Astrophys.\ J.\  {\bf 677}, L77 (2008).
   
\bibitem{Komatsu:2010fb}
  E.~Komatsu {\it et al.},  
  Astrophys.\ J.\ Suppl.\  {\bf 192}, 18 (2011).

\bibitem{Barkana:2004zy}
  R.~Barkana \& A.~Loeb,
  Astrophys.\ J.\  {\bf 624}, L65 (2005).

\bibitem{maoteg}
  Y.~Mao, {\it et al.},
  Phys.\ Rev.\  D {\bf 78}, 023529 (2008).

\bibitem{LCL}
A.~Lewis, {\it et al.}, Astrophys.\ J., {\bf 538}, 473 (2000).

\bibitem{Salopek:1990jq}
  D.~S.~Salopek \& J.~R.~Bond,
  PRD {\bf 42}, 3936 (1990).

\bibitem{Furlanetto:2004nh}
  S.~Furlanetto, {\it et al.},
  Astrophys.\ J.\  {\bf 613}, 1 (2004).

\bibitem{simfast21}
   M.~G.~Santos, {\it et al.},
  MNRAS {\bf 406}, 2421 (2010).

\bibitem{simfastlink}
http://www.simfast21.org

\bibitem{Matarrese:2000iz}
  S.~Matarrese, {\it et al.},
  Astrophys.\ J.\  {\bf 541}, 10 (2000).

\bibitem{mcquinn}
  M.~McQuinn, {\it et al.},
  Astrophys.\ J.\  {\bf 653}, 815 (2006).

\bibitem{Wyithe:2007if}
  S.~Wyithe \& M.~F.~Morales,
  arXiv:astro-ph/0703070.
  
\end{thebibliography}
\end{document}